\documentclass[a4paper]{article}
\usepackage{enumitem}
\usepackage{color}

\def\T{{\cal T}}

\usepackage{INTERSPEECH_v2}

\title{Improved I-vector-based Speaker Recognition for Utterances with Speaker Generated Non-speech sounds}
  \name{Sri Harsha Dumpala, Ashish Panda, Sunil Kumar Kopparapu}
\address{
TCS Innovation Labs-Mumbai, India}
 \email{\{d.harsha, ashish.panda, sunilkumar.kopparapu\}@tcs.com}

\begin{document}

\maketitle
\begin{abstract}
  Conversational speech not only contains several variants of neutral speech but is also prominently interlaced with several speaker
  generated non-speech sounds such as laughter and breath. A robust speaker recognition system should be capable of recognizing a 
  speaker irrespective of these variations in his speech. An understanding of whether the speaker-specific information represented by
  these variations is similar or not helps build a good speaker recognition system. In this paper, speaker variations captured by
  neutral speech of a speaker is analyzed by considering speech-laugh (a variant of neutral speech) and laughter (non-speech) sounds
  of the speaker. We study an i-vector-based speaker recognition system trained only on neutral speech and evaluate its performance
  on speech-laugh and laughter. Further, we analyze the effect of including laughter sounds during training of an i-vector-based
  speaker recognition system. Our experimental results show that the inclusion of laughter sounds during training seem to provide
  complementary speaker-specific information which results in an overall improved performance of the speaker recognition system,
  especially on the utterances with speech-laugh segments.
\end{abstract}

\section{Introduction}
\label{Sec:introduction}

The flexibility of the human speech production system not only allows production of several variants of neutral speech depending 
on the emotional and physical state of the speaker, but also allows production of several non-speech sounds such as laughter, cry,
cough, etc. These variations being produced by the speech production system of the same speaker might carry certain speaker-specific
information. However, it is not immediately clear if these variations in speech carry the {\em same} speaker-specific information
or not.

Speaker recognition refers to the task of identifying the speaker using speech as the only cue \cite{reynolds1995robust}. 
In recent years, speaker recognition systems have gained a significant improvement in performance making them more viable for 
commercial applications.
Current state-of-the-art speaker recognition systems employ i-vector-based approaches, where the feature sequence
representing the speech signal are characterized by low-dimensional fixed length vectors called identity-vectors (i-vectors in short)
\cite{dehak2011front}, \cite{saon2013speaker}, \cite{prince2007probabilistic}, \cite{sarkar2012study}. These i-vectors are
obtained by projecting the speech signal onto a subspace $\T$, referred to as ``total variability space'', which contains both speaker
and channel variabilities, simultaneously \cite{dehak2011front}.
I-vector-based approaches are well established as they have shown a significant improvement in speaker recognition
performances.
However, it is not immediately clear the relationship between different types of speech produced by a speaker 
on the performance of an i-vector-based speaker recognition system. More explicitly, there is a need to identify
if the {\em neutral} speech of a speaker is sufficient to represent the speaker-specific characteristics completely, 
irrespective of the variations in speech produced by the speaker. This information is essential for developing good and robust 
speaker recognition and diarization systems.

Only a few studies have analyzed the effect of considering different variations in speech, especially the
non-speech sounds such as
breath, whistle, scream, and breathy speech (a variant of neutral speech produced when breath co-occurs with speech) on the performance
of the speaker recognition systems trained using neutral speech, for example \cite{nandwana2014analysis}, \cite{nandwana2015new}, 
\cite{janicki2012impact} and more recently \cite{Harsha_IJCNN}. 
It is evident from earlier studies that the performance of the speaker recognition system trained using only neutral speech of the 
speaker degrades, if these variations in speech are a part of the testing phase of the system.
However, most of these studies considered traditional Gaussian mixture models with universal background
model (GMM-UBM) for their study but not the state-of-the-art i-vector-based systems. 
Also, these studies considered only non-speech sounds (such as breath, whistle etc.)
but not their speech co-occurring counterparts such as speech-laugh. 
In natural conversations, a significant part of the non-speech sounds co-occur with
speech to produce variants of neutral speech such as speech-laugh, breathy speech, etc., \cite{Harsha_IJCNN}, \cite{dumpala2016use},
\cite{hirose2010investigating}.
This emphasizes the need to analyze the robustness of the i-vector-based systems to such variations. 
Furthermore, to the best of our knowledge, none of the previous studies have analyzed the effect of
including these non-speech sounds in the enrollment phase of i-vector-based speaker recognition systems. 
Although, recent studies show that the inclusion of breath sounds in the training phase improves the
performance of GMM-UBM and deep neural network based speaker recognition systems on breathy speech \cite{Harsha_IJCNN},
it is to be systematically verified whether this observation holds true for other non-speech sounds, 
and even on i-vector-based systems.

The main objectives of this analysis are twofold. They are
\begin{enumerate}[noitemsep]
 \item To investigate whether the i-vectors extracted from the neutral speech of a speaker are robust to the various variations in
speech produced by the speaker.
\item To analyze the performance of i-vector-based speaker recognition systems when non-speech sounds along with neutral speech of the 
speaker are included in the enrollment phase of the systems.
\end{enumerate}

To achieve these objectives, speech-laugh (a variant of neutral speech) segments of the speaker are considered to evaluate the
performance of speaker recognition system developed using neutral speech. Further, laughter sounds collected from the speaker are 
included in the enrollment phase of the system. These systems are evaluated on both neutral speech and speech-laugh of the speaker
to find the presence of any complimentary speaker-specific information provided by laughter.
The significance of this analysis is evident from the fact that laughter is one of the most common non-speech sound which occurs
very frequently in natural conversations \cite{truong2005automatic}, and more than $50$\% of these laughter sounds happen to be speech-laughs 
\cite{nwokah1999integration}. 

The organization of the paper is as follows. Section \ref{Sec:Approach} explains the approach followed for analysis. 
Section \ref{Sec:ExpDetails} summarizes the dataset and the i-vector-based speaker recognition system considered. 
Experimental results are given in Section \ref{Sec:Results}.
Summary along with conclusions are given in Section \ref{Sec:Summary}

\begin{figure}[t]
\hspace{-1cm}
  \centering
  \includegraphics[width=1.02\linewidth]{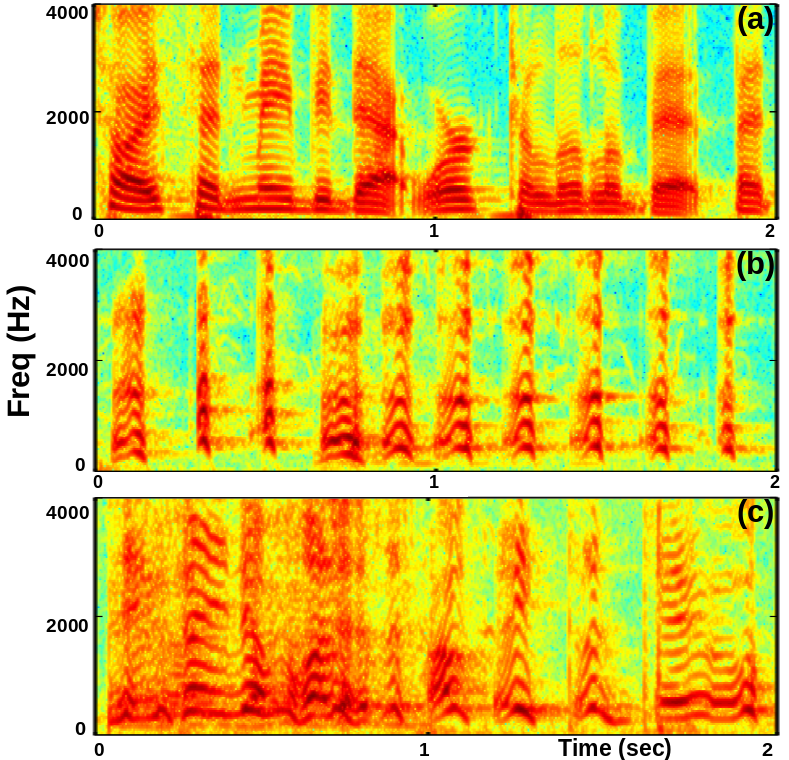}
  \caption{Spectrograms obtained for (a) neutral speech, (b) laughter and (c) speech-laugh, respectively.}
  \label{Fig:Spectrograms}
\end{figure}

\section{Background}
\label{Sec:Approach}
Analysis is performed by considering neutral speech, speech-laugh and laughter sounds produced by the speakers. Here, neutral speech 
refers to the normal/regular speech of the speaker. Laughter is a highly variable non-speech sound which is typically produced by a
series of sudden bursts of air through the vocal tract system \cite{wallace2007phonetics}.
Laughter sounds can be either voiced or unvoiced \cite{batliner2010laughter}, \cite{bachorowski2001acoustic}. In this analysis,
both, voiced and unvoiced laughter sounds of the speaker are considered. In conversational speech, laughter sounds frequently co-occur
with neutral speech to produce segments called speech-laugh \cite{trouvain2001phonetic}, \cite{menezes2006speech}. 
Speech-laugh is not produced by simply superimposing laughter on speech but its production involves a complex vocal configuration,
which exhibits characteristics of both laughter and neutral speech \cite{nwokah1999integration}, \cite{dumpala2014analysis}.
The spectral features obtained for laughter, speech-laugh and neutral speech form a continuum, with laughter exhibiting higher formant
frequencies (particularly, first formant frequency) followed by speech-laugh and then neutral speech \cite{menezes2006speech},
\cite{szameitat2011formant}. This can be observed from the spectrograms obtained for neutral speech, laughter and speech-laugh samples
of the same speaker (as shown in Figure \ref{Fig:Spectrograms}). This variation in formant frequencies, which might carry
speaker-specific information \cite{franco2016feature}, can effect the performance of the speaker recognition systems trained on neutral speech but tested on
speech-laugh and laughter segments. The effect of such variations on speaker recognition systems is analyzed in this study.

\begin{figure}[t]
\hspace{-0.9cm}
  \centering
  \includegraphics[width=8.8cm]{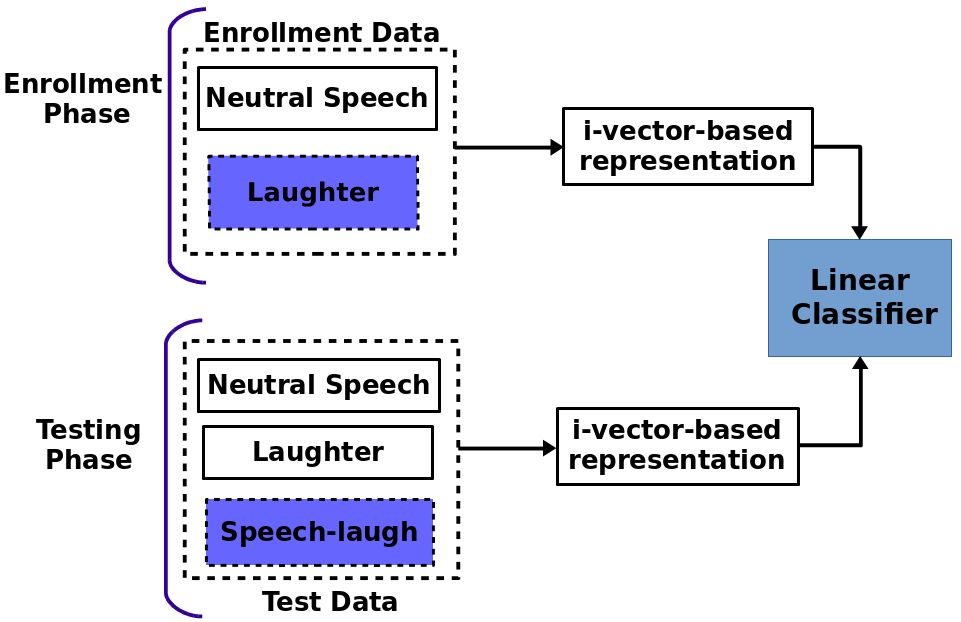}
  \caption{Block diagram of the approach followed for analysis.}
  \label{Fig:Approach}
\end{figure}

The approach followed for the analysis is depicted in Figure \ref{Fig:Approach}. It can be observed from Figure \ref{Fig:Approach} that
apart from neutral speech, laughter data collected from the speakers is also included in the enrollment phase of the speaker 
recognition system. For analysis, two different i-vector-based systems are developed, one using only neutral speech, and the other
is developed considering both, laughter and neutral speech of the speakers. The effect of including laughter in the enrollment
phase is analyzed by evaluating the performance of both systems when speech-laugh of the speakers (which is not available in the 
enrollment phase of either of the two systems) is provided as input apart from laughter and neutral speech of the speakers.

 \section{Experimental details}
 \label{Sec:ExpDetails}
For this analysis, GMM-UBM and i-vector statistics provided in the Voice biometry standardization (VBS) \cite{VBSToolkit} toolkit are
used.
In VBS toolkit, GMM-UBM with 2048 components was trained using NIST SRE 2004-2008 data and the T matrix required for i-vector extraction was trained
using Fisher English 
(Part $1$ and $2$), NIST SRE $2004-2008$, and Switchboard corpus (Phase $2$, Phase $3$, 
cellular part $1$ and cellular part $2$).
But standard speaker recognition corpus such as NIST SRE does not include speech transcripts, especially for non-speech sounds.
Hence, the enrollment and the test i-vectors used in this analysis are obtained using Buckeye corpus of
conversational speech \cite{pitt2007buckeye}.
Buckeye corpus of conversational speech, 
contains several hours of high quality recordings collected from $40$ speakers 
($20$ male and $20$ female).
This high quality conversational speech data was collected from the speakers 
in the form of informal interactions between the speaker and an interviewer. The 
corpus is phonetically labeled. Additionally, the laughter and speech-laugh segments 
produced by the speaker are separately labeled along with their timestamps. In this
work, we considered a subset of Buckeye corpus data collected from $30$ 
speakers ($15$ male and $15$ female), whose recordings have a significant amount of laughter and speech-laugh content.

\def\NS{{\sc Ns}}
\def\L{{\sc L}}
\def\SPL{{\sc Sl}}
\def\Tr{{\sc Es}}
\def\Ts{{\sc Ts}}
\def\Test{{\sc Ts}}
\def\Dataset{{\sc Dset}}
\def\enrollment{{enrollment}}
\def\training{{\enrollment}}
\def\System{{\sc System}}
\def\X{{$x$}}
\def\Y{{$y$}}
\def\Train{{\sc Es}}
\def\Tr{{\sc Es}}

\subsection{Data Organization}
For the purpose of analysis, we organized the Buckeye corpus into $7$ different datasets
as shown in Table \ref{Tab:DataDetails}. The speech of the speaker, in the corpus, 
is marked as neutral speech (\NS), laughter (\L) and speech-laugh (\SPL).
It can be observed from Table 
\ref{Tab:DataDetails} that enrollment sets (namely, \Tr1\ and \Tr2) of \Dataset1 (\NS) and \Dataset2 (\NS\ + \L) are used in \enrollment\ 
phase of the
speaker recognition system, whereas test sets of all datasets (i.e., \Ts1 through \Ts7) are used in the testing phase. 
The enrollment set in \Dataset2 (namely, \Train2) 
consists a total of $50$ utterances ($40$ \NS\ utterances and $10$ \L\ utterances) 
each of $2.5$ sec to $3$ sec in duration. But every utterance in \Test2 contains both 
\NS\ and \L. Similarly, every utterance in test set of 
\Dataset3, \Dataset4 and \Dataset7 (namely, \Test3, \Test4 and \Test7) contains the speech types as specified in Table \ref{Tab:DataDetails}.
It is to be noted that \SPL\ is not used in the \enrollment\ phase but only in the testing 
phase to analyze the effect of considering laughter sounds in the enrollment phase of speaker recognition systems.
\begin{table*}[t]
\renewcommand{\arraystretch}{1.1}
 \caption{{Dataset organization details (Enrollment (\Train) and Test (\Test) refer to Enrollment set and Test set, respectively).}}
 \label{Tab:DataDetails}
 \centering 
\begin{tabular}{|c|c|c|c|c|} \hline 
& & \multicolumn{2}{c|}{\# Utterances/speaker} & \\ \cline{3-4} 
Dataset & Contents & Enrollment (\Train) & Test (\Test) & Duration (sec) \\ \hline
 \Dataset1 & \multicolumn{1}{l|}{Neutral speech (\NS)} &\Train1 = 50 &\Test1 = 25& 2.5-3 \\ \hline 
\Dataset2 & \multicolumn{1}{l|}{Neutral speech and laughter (\NS\ + \L)} &\Tr2 = 50 &\Ts2 = 15 & 2.5-3 \\ \hline 
\Dataset3 & \multicolumn{1}{l|}{Neutral speech and speech-laugh (\NS\ + \SPL)} & - &\Ts3 = 15 & 2.5-3 \\ \hline 
\Dataset4 & \multicolumn{1}{l|}{Neutral speech, laughter and speech-laugh (\NS\ + \L\ + \SPL)} & - &\Ts4 = 15 & 2.5-3 \\ \hline 
\Dataset5 & \multicolumn{1}{l|}{Laughter (\L)} & - &\Ts5 = 10 & 2.5-3 \\ \hline 
\Dataset6 & \multicolumn{1}{l|}{Speech-laugh (\SPL)} & - &\Ts6 = 10 & 2.5-3 \\ \hline 
 \Dataset7 & \multicolumn{1}{l|}{Laughter and speech-laugh (\L\ + \SPL)} & - &\Ts7 = 10 & 2.5-3 \\ \hline

 \end{tabular} 
 \end{table*}
 
 \subsection{System description}

\begin{figure}
 \hspace{-0.4cm}
 \includegraphics[width=1.13\linewidth,height=7cm,keepaspectratio]{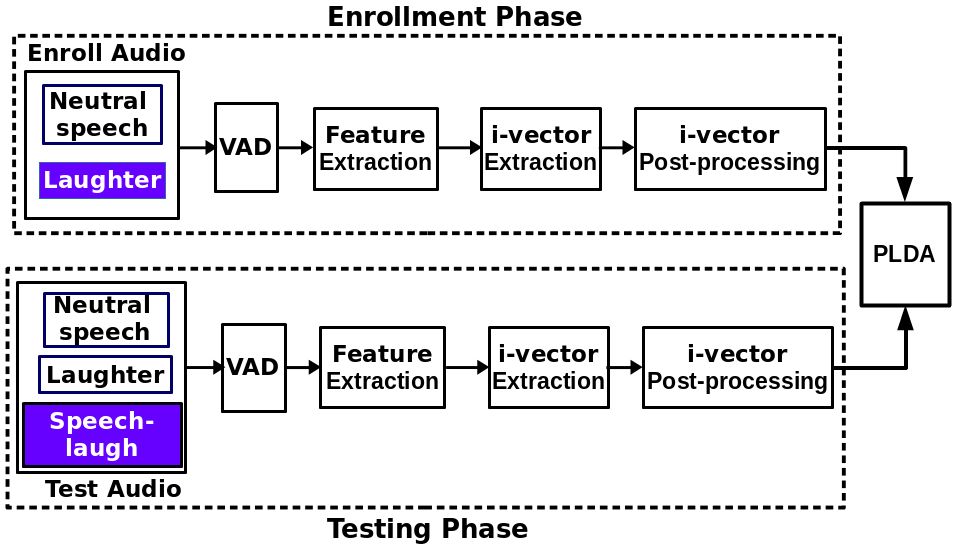} \caption{Block diagram of 
 i-vector-based speaker recognition system implementation.} \label{Fig:BlockDiag} 
  \vspace{-0.3cm}
 \end{figure}

 The i-vector-based speaker recognition systems considered for analysis are 
implemented using the VBS \cite{VBSToolkit} toolkit. 
Figure \ref{Fig:BlockDiag} shows the schematic of the 
system implementation using VBS and consists of the audio, voice activity detection (VAD), feature extraction, i-vector extraction and
post processing.
We use probabilistic linear discriminant analysis (PLDA) as the metric to measure the performance of the speaker recognition system.
We describe these blocks in more detail \cite{VBSToolkit}.

\textbf {Audio:} The audio data considered in the \training\ phase (see Enroll Audio in Figure \ref{Fig:BlockDiag}) consists of
\NS\ and \L\ sounds of each speaker. 
Whereas the audio data in the testing phase (see Test Audio in Figure \ref{Fig:BlockDiag}) consists of \NS, \L\ and \SPL.
All the audio samples are down sampled to $8$ kHz and are in $16$-bit PCM format as required by the VBS.

\textbf{Voice activity detection (VAD):} VAD is used prior to feature extraction 
to remove the silence and low signal-to-noise ratio (SNR) regions in the audio sample. In this analysis, 
VAD is performed using the VOICEBOX toolkit \cite{brookes1997voicebox}.


\textbf{Feature extraction:} Regions of the audio signal retained after VAD are 
represented using mel-frequency cepstral coefficients (MFCCs). MFCCs are extracted 
using $25$ msec Hamming window with $10$ msec forward shift. 
MFCCs are computed by using $24$ mel filter banks and limiting the bandwidth to 
frequency in the $125$ Hz - $3800$ Hz range. 
Every frame is represented using $20$ coefficients 
(first $19$ MFCCs along with the $0^{th}$ coefficient). This $20$-dimensional 
feature vector is mean and variance normalized using $3$ sec sliding window. Subsequently, 
the delta and the double delta coefficients are computed to form a 
$60$-dimensional feature vector to represent each frame. 

\textbf{i-vector extraction:} To obtain a low-dimensional fixed-length 
i-vector-based representation of the sequence of feature vectors, the GMM-UBM and 
the i-vector statistics (total variability space (T)) are necessary. In this 
analysis, the GMM-UBM-based i-vectors are extracted using the GMM-UBM and T 
matrix statistics released by VBS. The gender-independent 
universal background model (UBM) with $2048$ components was trained using
NIST SRE $2004-2008$ data ($\approx 1156.03$ hours of data). The 
total variability space 'T' of $600$-dimension was trained using Fisher English 
(Part $1$ and $2$), NIST SRE $2004-2008$, Switchboard (Phase $2$, Phase $3$, 
cellular part $1$ and cellular part $2$) which totals to $9010.23$ hours of data.

\textbf{i-vector post-processing:}
The i-vectors of $600$-dimensions obtained for each audio sample are reduced to $200$-dimensions using linear discriminant analysis
(LDA) \cite{dehak2011front}, \cite{garcia2011analysis}.
Then these i-vectors are further normalized using within-class covariance matrix 
\cite{garcia2011analysis}. Both, LDA and within-class covariance matrix are 
provided by VBS and are trained on the same data that is used for `T' matrix 
\training. In this analysis, the speaker templates (namely, i-vectors corresponding to 
each speaker) are generated separately for the two considered cases (namely, \NS\ and 
\NS\ + \L) during the \enrollment\ phase, and an i-vector is obtained for each audio 
sample during the test phase.

\textbf{PLDA:} To compare the \enrollment\ i-vectors to the test i-vectors for 
speaker recognition, PLDA is used 
\cite{prince2007probabilistic}, \cite{kenny2010bayesian}. PLDA is a special case of 
joint factor analysis (JFA) with single Gaussian component, but is used in the 
i-vector space. Given a pair of i-vectors, PLDA computes the log-likelihood score 
for the same-speaker and the different-speaker hypothesis 
\cite{burget2011discriminatively}. This score is used to evaluate the 
speaker recognition system.

 \begin{table}[h] 
\renewcommand{\arraystretch}{1.1} 
\caption{EER (in \%) obtained for the systems on different datasets (\NS, \L\ and \SPL\ refers to Neutral speech, laughter
and speech-laugh, respectively and, \System1 (baseline) is trained on \NS\ and \System2 (proposed) is trained on \NS\ + \L).} 

\label{Tab:EER values} 
\centering 
\begin{tabular}{|l|c|c|}
  \hline Test-set & \System1 (\Tr1) & \System2 (\Tr2) \\
  & (Baseline) & (Proposed) \\ \hline 
\Ts1 (\NS) & 2.69 & 2.75 \\ \hline 
\Ts2 (\NS\ + \L) & 9.03 & 4.32\\ \hline 
\Ts3 (\NS\ + \SPL) & 6.37 & 4.71 \\ \hline 
\Ts4 (\NS\ + \L\ + \SPL) & 8.69 & 5.83\\ \hline 
\Ts5 (\L) & 27.63 & 13.94 \\ \hline 
\Ts6 (\SPL) & 13.67 & 9.92\\ \hline 
\Ts7 (\L\ + \SPL) & 17.97 & 12.02 \\ 
  \hline
  
 \end{tabular} \end{table}

 \section{Experimental results} \label{Sec:Results} The performance of the 
i-vector-based speaker recognition systems considered is evaluated in terms of 
equal error rate (EER), where lower the EER value, better is the performance of the 
system. The EER values (in \%) obtained for the two considered systems, namely, \System1 
(trained on \Tr1 i.e., \NS) and \System2 (trained on \Tr2 i.e., \NS\ + \L), when tested on all the test datasets 
(\Ts1 through \Ts7) are shown in Table \ref{Tab:EER values} (refer to Table \ref{Tab:DataDetails} for dataset details). 
As observed from Table \ref{Tab:EER values}
 \begin{itemize}
\item Both \System1 and \System2 perform equally well when tested on 
\Ts1 with EER of $2.69\%$ and $2.75\%$, respectively. This shows that the inclusion of laughter sounds in enrollment phase have
little (no effect) on the performance of the speaker recognition system, when test utterances consist of only neutral speech of the
speakers.

\item A higher degradation in performance is observed for \System1 compared to 
\System2, when laughter (\L) is included in the test set (\Ts2, \Ts5). This shows that the 
i-vector-based speaker representation obtained from neutral speech (\NS) of speaker is 
different from that of laughter, signifying the variation in speaker-specific 
information exhibited by neutral speech and laughter. 

\item \System2 performs better 
than \System1, when speech-laugh (which is not present in either \Tr1 or \Tr2) is 
present in the test set (\Ts3, \Ts4, \Ts6, \Ts7). This seems to indicate 
that the laughter sounds 
provide complementary speaker-specific information when compared with neutral 
speech, which results in better speaker recognition in case of speech-laugh.
For instance when tested on \Ts6 (\SPL), \System2 (EER=$9.92\%$) achieved a relative improvement of $27\%$ (in terms of EER) compared to
\System1 (EER=$13.67\%$). 

\item The importance of considering laughter in 
\enrollment\ phase is evident from the performance of \System2 on \Ts4 (EER=$5.83\%$), which gained a relative improvement of around $33\%$ 
(in EER) compared to \System1 (EER=$8.69\%$), where \Ts4 (\NS\ + \L\ + \SPL) is the most 
common format in conversational speech. \end{itemize}

\begin{figure}[t]
  \centering
  \includegraphics[height=9.2cm]{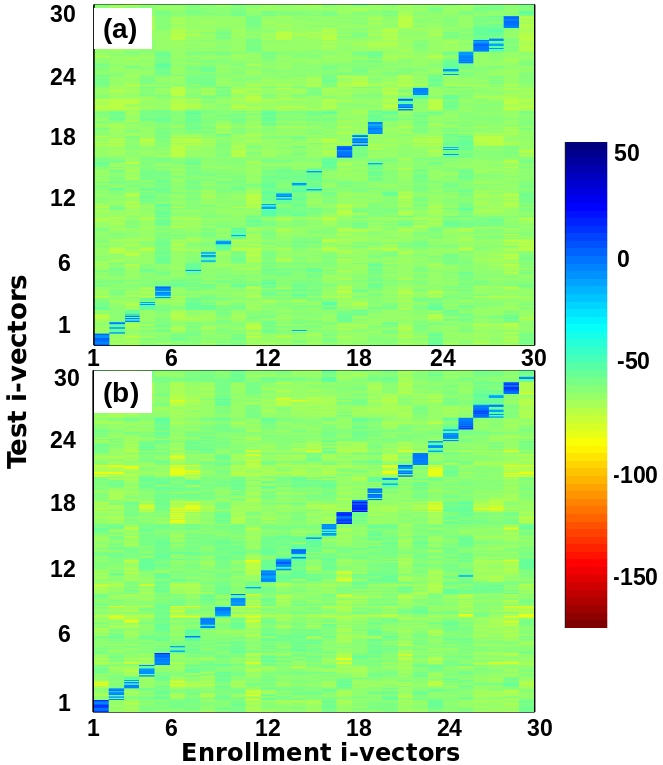}
  \caption{Figure shows the PLDA scores obtained for (a) \System 1 (trained on \Tr1) and (b) \System 2 (trained on
  \Tr2), when tested on \Ts6. (Note: above Figure is best viewed in color)}
  \label{Fig:iscores}
  \vspace{-1em}
\end{figure}

 Furthermore, the PLDA scores (log likelihood values) obtained between 
\enrollment\ i-vectors and test i-vectors are analyzed to justify the EER values 
shown in Table \ref{Tab:EER values}. Figure \ref{Fig:iscores} shows the PLDA scores 
obtained between enrollment i-vectors (i-vectors obtained from (a) \Tr1 and (b) \Tr2) 
and test i-vectors (i-vectors obtained from \Ts6). In this plot, each value 
represents the PLDA score obtained between an \enrollment\ i-vector (along \X-axis) 
and a test i-vector (along \Y-axis). For instance, the value at position $(n, m)$
refers to the PLDA score obtained between $n^{th}$ \enrollment\ i-vector (corresponds 
to $n^{th}$ \enrollment\ speaker) and $m^{th}$ test i-vector (corresponds to $m^{th}$ 
speaker test utterance). A better performing system (lower EER) will exhibit higher 
values along the diagonal (namely, $n = m$) and lower values at other locations ($n 
\neq m$). It can be observed from Figure \ref{Fig:iscores} that the PLDA scores 
along the diagonal are higher for \System2 (\enrollment\ i-vectors obtained using \Tr2) 
compared to the corresponding values obtained for \System1 (\enrollment\ i-vectors 
obtained using \Tr1), when tested on \Ts6 (\SPL). Also, the difference between the 
PLDA scores along the diagonal, and at other locations is higher for \System2 
compared to that of \System1. This explains the lower EER values obtained by \System2 
for utterances with speech-laugh (\Ts3, \Ts4, \Ts6 and \Ts7) compared to the 
corresponding EER values obtained for \System1.


 \section{Summary and conclusions}
 \label{Sec:Summary} 

Natural conversations between people have significant amount of non-speech events, such as laughter, speech-laugh, interlaced with neural
speech. A practical speaker recognition system needs to be able to recognize a speaker in these scenarios.
We used the i-vector to represent the speaker characteristic.
In this paper, the variation 
in speaker-specific information captured by different variants of neutral speech 
and non-speech sounds, produced by the same speaker, was experimentally 
evaluated by considering 
neutral speech, speech-laugh and laughter sounds. 
This analysis is performed
on an i-vector-based speaker recognition system.
Experimental results show that the i-vector-based speaker representation obtained 
using neutral speech of a speaker differs from the speaker representation obtained 
using laughter sounds and speech-laugh segments produced by the same speaker. 
Further, the complementary speaker-specific information provided by laughter sounds 
of the speaker will help in improving the performance of i-vector-based speaker 
recognition systems not only for utterances containing laughter sounds, but also in 
the case of speech-laugh.
\bibliographystyle{IEEEtran}

\bibliography{references}

\begin{thebibliography}{10}
\providecommand{\url}[1]{#1}
\csname url@samestyle\endcsname
\providecommand{\newblock}{\relax}
\providecommand{\bibinfo}[2]{#2}
\providecommand{\BIBentrySTDinterwordspacing}{\spaceskip=0pt\relax}
\providecommand{\BIBentryALTinterwordstretchfactor}{4}
\providecommand{\BIBentryALTinterwordspacing}{\spaceskip=\fontdimen2\font plus
\BIBentryALTinterwordstretchfactor\fontdimen3\font minus
  \fontdimen4\font\relax}
\providecommand{\BIBforeignlanguage}[2]{{%
\expandafter\ifx\csname l@#1\endcsname\relax
\typeout{** WARNING: IEEEtran.bst: No hyphenation pattern has been}%
\typeout{** loaded for the language `#1'. Using the pattern for}%
\typeout{** the default language instead.}%
\else
\language=\csname l@#1\endcsname
\fi
#2}}
\providecommand{\BIBdecl}{\relax}
\BIBdecl

\bibitem{reynolds1995robust}
D.~A. Reynolds and R.~C. Rose, ``Robust text-independent speaker identification
  using gaussian mixture speaker models,'' \emph{IEEE transactions on speech
  and audio processing}, vol.~3, no.~1, pp. 72--83, 1995.

\bibitem{dehak2011front}
N.~Dehak, P.~J. Kenny, R.~Dehak, P.~Dumouchel, and P.~Ouellet, ``Front-end
  factor analysis for speaker verification,'' \emph{IEEE Transactions on Audio,
  Speech, and Language Processing}, vol.~19, no.~4, pp. 788--798, 2011.

\bibitem{saon2013speaker}
G.~Saon, H.~Soltau, D.~Nahamoo, and M.~Picheny, ``Speaker adaptation of neural
  network acoustic models using i-vectors.'' in \emph{ASRU}, 2013, pp. 55--59.

\bibitem{prince2007probabilistic}
S.~J. Prince and J.~H. Elder, ``Probabilistic linear discriminant analysis for
  inferences about identity,'' in \emph{Computer Vision, 2007. ICCV 2007. IEEE
  11th International Conference on}.\hskip 1em plus 0.5em minus 0.4em\relax
  IEEE, 2007, pp. 1--8.

\bibitem{sarkar2012study}
A.~K. Sarkar, D.~Matrouf, P.-M. Bousquet, and J.-F. Bonastre, ``Study of the
  effect of i-vector modeling on short and mismatch utterance duration for
  speaker verification.'' in \emph{Interspeech}, 2012, pp. 2662--2665.

\bibitem{nandwana2014analysis}
M.~K. Nandwana and J.~H. Hansen, ``Analysis and identification of human scream:
  implications for speaker recognition.'' in \emph{INTERSPEECH}, 2014, pp.
  2253--2257.

\bibitem{nandwana2015new}
M.~K. Nandwana, H.~Boril, and J.~H. Hansen, ``A new front-end for
  classification of non-speech sounds: a study on human whistle.'' in
  \emph{INTERSPEECH}, 2015, pp. 1982--1986.

\bibitem{janicki2012impact}
A.~Janicki, ``On the impact of non-speech sounds on speaker recognition,'' in
  \emph{International Conference on Text, Speech and Dialogue}.\hskip 1em plus
  0.5em minus 0.4em\relax Springer, 2012, pp. 566--572.

\bibitem{Harsha_IJCNN}
S.~H. Dumpala and S.~K. Kopparapu, ``Improved speaker recognition system for
  stressed speech using deep neural networks,'' in \emph{IEEE International
  Joint Conference on Neural Networks (IJCNN)}, 2017, accepted for publication.

\bibitem{dumpala2016use}
S.~H. Dumpala, P.~Gangamohan, S.~V. Gangashetty, and B.~Yegnanarayana, ``Use of
  vowels in discriminating speech-laugh from laughter and neutral speech,''
  \emph{Interspeech 2016}, pp. 1437--1441, 2016.

\bibitem{hirose2010investigating}
H.~Hirose, ``Investigating the physiology of laryngeal structures,'' \emph{The
  handbook of phonetic sciences}, pp. 130--52, 2010.

\bibitem{truong2005automatic}
K.~P. Truong and D.~A. Van~Leeuwen, ``Automatic detection of laughter.'' in
  \emph{INTERSPEECH}, 2005, pp. 485--488.

\bibitem{nwokah1999integration}
E.~E. Nwokah, H.-C. Hsu, P.~Davies, and A.~Fogel, ``The integration of laughter
  and speech in vocal communicationa dynamic systems perspective,''
  \emph{Journal of Speech, Language, and Hearing Research}, vol.~42, no.~4, pp.
  880--894, 1999.

\bibitem{wallace2007phonetics}
C.~Wallace, ``The phonetics of laughter--a linguistic approach,'' in
  \emph{Interdisciplinary Workshop on the Phonetics of Laughter}, 2007, pp.
  4--5.

\bibitem{batliner2010laughter}
A.~Batliner, S.~Steidl, F.~Eyben, and B.~Schuller, ``On laughter and speech
  laugh, based on observations of child-robot interaction,'' \emph{The
  phonetics of laughing, trends in linguistics. de Gruyter, Berlin, to appear},
  2010.

\bibitem{bachorowski2001acoustic}
J.-A. Bachorowski, M.~J. Smoski, and M.~J. Owren, ``The acoustic features of
  human laughter,'' \emph{The Journal of the Acoustical Society of America},
  vol. 110, no.~3, pp. 1581--1597, 2001.

\bibitem{trouvain2001phonetic}
J.~Trouvain, ``Phonetic aspects of “speech-laughs”,'' in \emph{Oralit{\'e}
  et Gestualit{\'e}: Actes du colloque ORAGE, Aix-en-Provence. Paris:
  L’Harmattan}, 2001, pp. 634--639.

\bibitem{menezes2006speech}
C.~Menezes and Y.~Igarashi, ``The speech laugh spectrum,'' \emph{Proc. Speech
  Production, Brazil}, pp. 157--164, 2006.

\bibitem{dumpala2014analysis}
S.~H. Dumpala, K.~V. Sridaran, S.~V. Gangashetty, and B.~Yegnanarayana,
  ``Analysis of laughter and speech-laugh signals using excitation source
  information,'' in \emph{Acoustics, Speech and Signal Processing (ICASSP),
  2014 IEEE International Conference on}.\hskip 1em plus 0.5em minus
  0.4em\relax IEEE, 2014, pp. 975--979.

\bibitem{szameitat2011formant}
D.~P. Szameitat, C.~J. Darwin, A.~J. Szameitat, D.~Wildgruber, and K.~Alter,
  ``Formant characteristics of human laughter,'' \emph{Journal of voice},
  vol.~25, no.~1, pp. 32--37, 2011.

\bibitem{franco2016feature}
J.~Franco-Pedroso and J.~Gonzalez-Rodriguez, ``Feature-based likelihood ratios
  for speaker recognition from linguistically-constrained formant-based
  i-vectors,'' \emph{Odyssey 2016}, pp. 237--244, 2016.

\bibitem{VBSToolkit}
``Voice biometry standardization {(VBS)} initiative,''
  \url{http://voicebiometry.org/}, 2015.

\bibitem{pitt2007buckeye}
M.~A. Pitt, L.~Dilley, K.~Johnson, S.~Kiesling, W.~Raymond, E.~Hume, and
  E.~Fosler-Lussier, ``Buckeye corpus of conversational speech (2nd release),''
  \emph{Columbus, OH: Department of Psychology, Ohio State University}, 2007.

\bibitem{brookes1997voicebox}
M.~Brookes \emph{et~al.}, ``Voicebox: Speech processing toolbox for matlab,''
  \emph{Software, available [Mar. 2011] from www. ee. ic. ac.
  uk/hp/staff/dmb/voicebox/voicebox. html}, vol.~47, 1997.

\bibitem{garcia2011analysis}
D.~Garcia-Romero and C.~Y. Espy-Wilson, ``Analysis of i-vector length
  normalization in speaker recognition systems.'' in \emph{Interspeech}, vol.
  2011, 2011, pp. 249--252.

\bibitem{kenny2010bayesian}
P.~Kenny, ``Bayesian speaker verification with heavy-tailed priors.'' in
  \emph{Odyssey}, 2010, p.~14.

\bibitem{burget2011discriminatively}
L.~Burget, O.~Plchot, S.~Cumani, O.~Glembek, P.~Mat{\v{e}}jka, and
  N.~Br{\"u}mmer, ``Discriminatively trained probabilistic linear discriminant
  analysis for speaker verification,'' in \emph{Acoustics, Speech and Signal
  Processing (ICASSP), 2011 IEEE International Conference on}.\hskip 1em plus
  0.5em minus 0.4em\relax IEEE, 2011, pp. 4832--4835.

\end{thebibliography}

\end{document}